\documentclass[conference]{IEEEtran}
\usepackage{cite, wrapfig, amsmath, amsfonts, eqparbox, array, stfloats, url, xcolor, tabularx, colortbl, booktabs, indentfirst, booktabs,amssymb,amsthm,amscd,dsfont,mathrsfs,comment,url,paralist,caption}
\usepackage[pdftex]{graphicx} 
\usepackage{rotating}
\usepackage[english]{babel}
\usepackage[tight,footnotesize]{subfigure}
\usepackage[euler]{textgreek}
\usepackage{float}
\restylefloat{table}
\DeclareGraphicsExtensions{.pdf,.jpeg,.png}
\theoremstyle{plain}
\theoremstyle{plain}
\theoremstyle{plain}
\theoremstyle{plain}
\theoremstyle{plain}
\newtheorem{fact}{Fact}
\theoremstyle{plain}
\theoremstyle{plain}
\theoremstyle{plain}
\theoremstyle{plain}
\theoremstyle{remark}
\newtheorem{remark}{Remark}
\theoremstyle{discussion}
\theoremstyle{plain}

\definecolor{light-gray}{gray}{0.85}

\begin{document}
\title{A 2.48Gb/s QC-LDPC Decoder Implementation on the NI USRP-2953R} 
\author
{\IEEEauthorblockN{Swapnil Mhaske\IEEEauthorrefmark{1}, David Uliana\IEEEauthorrefmark{2}, Hojin Kee\IEEEauthorrefmark{2}, Tai Ly\IEEEauthorrefmark{2}, Ahsan Aziz\IEEEauthorrefmark{2}, Predrag Spasojevic\IEEEauthorrefmark{1}}
\IEEEauthorblockA{\IEEEauthorrefmark{1}Wireless Information Network Laboratory, Rutgers University, New Brunswick, NJ, USA \\
Email:\{swapnil.mhaske@rutgers.edu, spasojev@winlab.rutgers.edu\}}
\IEEEauthorblockA{\IEEEauthorrefmark{2}National Instruments Corporation, Austin, TX, USA \\
Email:\{david.uliana, hojin.kee, tai.ly, ahsan.aziz\}@ni.com}}
\maketitle

\begin{abstract}
The increasing data rates expected to be of the order of Gb/s for future wireless systems directly impact the throughput requirements of the modulation and coding subsystems of the physical layer. In an effort to design a suitable channel coding solution for 5G wireless systems, in this brief we present a massively-parallel 2.48Gb/s Quasi-Cyclic Low-Density Parity-Check (QC-LDPC) decoder implementation operating at 200MHz on the \emph{NI USRP-2953R}, on a single FPGA. The algorithmic innovations leading to an architecture-aware design, central to this work, are presented in \cite{arch_main}. The high-level description of the entire massively-parallel decoder was translated to a Hardware Description Language (HDL), namely VHDL, using the algorithmic compiler \cite{algcmp} in the \emph{National Instruments \emph{LabVIEW\texttrademark} \,Communication System Design Suite (CSDS\texttrademark)} in approximately 2 minutes. This implementation not only demonstrates the scalability of our initial work in \cite{arch_main} but also, the rapid prototyping capability of the \emph{\emph{LabVIEW\texttrademark} CSDS\texttrademark} tools. As per our knowledge, at the time of writing this paper, this is the fastest implementation of a standard compliant QC-LDPC decoder on a USRP using an algorithmic compiler.
\end{abstract}

\begin{IEEEkeywords}
5G, mm-wave, SDR, USRP, QC-LDPC, layered decoding.
\end{IEEEkeywords}

\section{Introduction}
\label{sec:intro}
Wireless data traffic is expected to increase by a 1000 fold \cite{5g_nsn_raaf} by  the year 2020 with more than 50 billion devices connected to these wireless networks with peak data rates upto ten Gb/s \cite{5g_nsn_cudak}. To address these challenges, the next generation of wireless cellular technology being envisioned and researched today is collectively termed as Beyond-4G (B-4G) and 5G. However, the envisioned operation of 5G systems in the mm-wave (30-300GHz) spectrum comes with  challenges such as, reliance on line of sight (LOS) communication, short range of communication, significantly increased shadowing loss and rapid fading in time, necessitating techniques such as large antenna arrays and rapidly adaptive beamsteering. From a physical layer perspective, the processing budget (especially time)  available to the channel encoder and decoder will further decrease (relative to current generation systems such as 4G LTE). \\
\indent With this in mind, in our ongoing research we focus on high-throughput and low-latency error control coding solutions (primarily based on Low-Density Parity-Check (LDPC) \cite{ldpc_gallager} family of codes) specially suited to 5G mm-wave systems. At the time of writing this paper, a detailed progress report focusing on the algorithmic innovations for high-throughput and a subsequent case study leading to a 608Mb/s (at 260MHz) standard compliant Quasi-Cyclic (QC) LDPC decoder is presented in \cite{arch_main} and \cite{ht_arch}. To adapt to the evolving specifications for 5G technology, implementations for our ongoing research must be reconfigurable and scalable, and must exhibit state-of-the-art performance, hence we choose the FPGA approach to developing hardware instead of the ASIC approach. \\
The Universal Software Radio Peripheral (USRP) is a widely used Software Defined Radio (SDR) system that is a flexible and an affordable transceiver with the potential to turn a standard host (such as a PC) into a powerful wireless prototyping system. The availability of state-of-the-art, highly reconfigurable hardware platforms (such as the FPGA) on the USRP has opened up a huge space for implementing theoretical algorithms at high-speeds, crucial for systems such as those required by 5G wireless. \\
\indent In this brief we present an application of the work in \cite{arch_main}, a 2.48Gb/s FPGA-based QC-LDPC decoder implemented on the \emph{NI USRP-2953R} (which has the \emph{Xilinx Kintex7 (410t)} FPGA) using the \emph{FPGA IP} compiler in \emph{LabVIEW\texttrademark \,CSDS\texttrademark}. Massive-parallelization was accomplished by employing 6 decoder cores in parallel without any modification at the HDL level. This compiler translated the entire high-level description of the parallelization (done in a graphical algorithmic dataflow language) to VHDL and further generated an optimized hardware implementation from the algorithmic description. The main contributions of this work are: (1) demonstration of the scalability of our decoder architecture in \cite{arch_main} (2) the ability of the \emph{LabVIEW\texttrademark \,CSDS\texttrademark} tools to rapidly prototype high-level algorithmic description onto FPGA hardware. \\
\indent The remainder of this paper is organized as follows. Section \ref{sec:qcldpc} outlines the construction of QC-LDPC codes and the decoding algorithm used for the implementation. A brief overview of the techniques leading to the software-pipelined decoder core in \cite{arch_main} is given in Section \ref{sec:swppl}. The process of implementing the 2.48Gb/s decoder and the performance results are detailed in Sections \ref{sec:sixcores} and \ref{sec:results} respectively, and finally we conclude with Section \ref{sec:conc}.

\section{Quasi-Cyclic LDPC Codes and Decoding}
\label{sec:qcldpc}
Mathematically, given $k$ message bits, an LDPC code is a null-space of its $m \times n$ PCM $\mathbf{H}$, where $m$ denotes the number of parity-check equations or parity-bits and $n \, (=k+m)$ denotes the number of variable nodes or code bits \cite{ecc_shulin}. In the Tanner graph representation (due to Tanner \cite{ldpc_tanner}), $\mathbf{H}$ is the incidence matrix of a bipartite graph comprising of two sets: the check node (CN) set of $m$ parity-check equations and the variable node (VN) set of $n$ variable or bit nodes; the $i^{th}$ CN is connected to the $j^{th}$ VN if $\mathbf{H}(i,j)=1$, $1 \leq i \leq m$ and $1 \leq j \leq n$.

QC-LDPC codes are represented by an $m_b  \times n_b$ \emph{base} matrix $\mathbf{H}_b$ which comprises of cyclically right-shifted identity and zero submatrices both of size $z \times z$ where, $z \in \mathbb{Z^+}, 1 \leq i_b \leq m_b$ and $1 \leq j_b \leq n_b$, the shift value,$s = \mathbf{H}_b(i_b,j_b) \in \mathcal{S} = \{-1\} \cup \{0, \ldots z-1\}$
The PCM matrix $\mathbf{H}$ is obtained by \emph{expanding} $\mathbf{H}_b$ using the mapping,
\begin{align*}
s \longrightarrow \left\{
  \begin{array}{lr}
  \mathbf{I}_s, &s \in \mathcal{S} \backslash \{-1\}\\
  \mathbf{0}, &s \in \{-1\}
  \end{array}
\right.
\end{align*}
where, $\mathbf{I}_s$ is an identity matrix of size $z$ which is cyclically right-shifted by $s=\mathbf{H}_b(i_b,j_b)$ and $\mathbf{0}$ is the all-zero matrix of size $z \times z$. Owing to this structure provided by QC-LDPC codes, the decoding of these codes becomes much simpler in hardware (mainly due to the simplified interconnect complexity) compared to unstructured LDPC codes. We believe that the family of structured LDPC codes are highly likely candidates for 5G systems. Thus, to demonstrate the initial phase of our FPGA decoder architecture \cite{arch_main}, \cite{ht_arch}, we provide a case study based on the QC-LDPC code specified in the \emph{IEEE 802.11n (2012)} standard \cite{std80211n2012}, the throughput of which well surpasses the requirement of the standard. \\
\indent LDPC codes can be decoded using message passing (MP) or belief propagation (BP) \cite{ldpc_gallager,spa_factorgraphs} on the bipartite Tanner graph where, the CNs and VNs communicate with each other, successively passing revised estimates of the log-likelihood ratio (LLR) associated, in every decoding iteration. The decoder in \cite{arch_main} employs the efficient decoding algorithm in \cite{serialmp_litsyn}, with pipelined processing of layers based on the row-layered decoding technique in \cite{laydec}. A stepwise description of the version of the algorithm we have employed is given in \cite{arch_main}. 

\section{Software Pipelined Decoder Architecture}
\label{sec:swppl}
\begin{figure*}
\centering
\includegraphics[width=\linewidth]{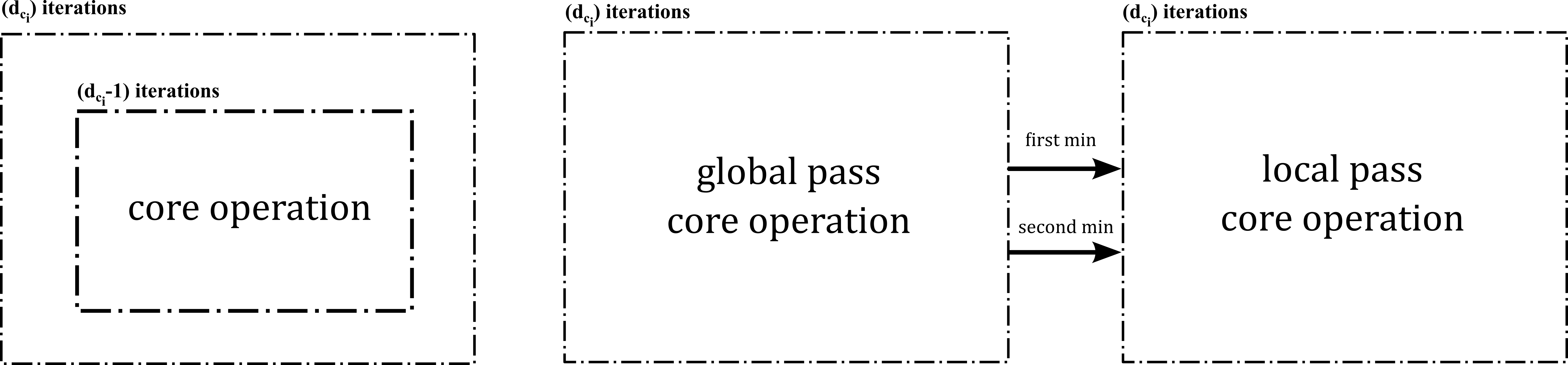}
\caption{Illustration of software pipelining of layers for the case study. (a) Layer-level processing view for the Baseline architecture showing the cascade of the GNPU and LNPU for-loops (b) Layer-level processing view for the Pipelined architecture showing the two for-loops operating in parallel for half of a decoding iteration. Note that, GNPU (LNPU) array refers to a collection of $z$ GNPUs (LNPUs) which operate in parallel so as to exploit the structure provided by QC-LDPC codes as shown in \cite{arch_main}.}
\label{fig:swppl}
\end{figure*}

Without loss of generality, in \cite{arch_main} we have presented several strategies to achieve high-throughput for the decoder architecture. To understand how software-pipelining was accomplished for a single core (amongst the 6 parallel cores in this implementation) and for the sake of continuity and completeness, we provide an overview of the layer decoder architecture from \cite{arch_main} below. \\
\indent From the perspective of CN processing, two or more CNs can be processed at the same time (i.e. they are independent of each other) if they do not have one or more VNs (code bits) in common. In terms of $\mathbf{H}$, an arbitrary subset of rows can be processed at the same time provided that, no two or more rows have a $1$ in the same column of $\mathbf{H}$. This subset of rows is termed as a \emph{row-layer} (hereafter referred to as a \emph{layer}). In other words, given a set $\mathcal{L}=\{L_1, L_2, \ldots,L_I\}$ of $I$ layers in $\mathbf{H}$, $\forall u \in \{1,2,\ldots,I\}$ and $\forall i, i^\prime \in L_u$, then, $\mathcal{N}(i) \cap \mathcal{N}(i^\prime)=\phi.$ Owing to the structure of QC-LDPC codes, $|L_u|=z$. From the VN or column perspective, $|L_u|=z$, $\forall u = \{1,2,\ldots,I\}$ implies that, the columns of $\mathbf{H}$ are also divided into subsets of size $z$ (hereafter referred to as \emph{block columns}) given by the set $\mathcal{B}=\{B_1,B_2,\ldots,B_J\}$, $J=\frac{n}{z}=n_b$. Since, VNs in a block column may participate in CN equations across several layers, we further divide the block columns into \emph{blocks}, where a block is the intersection of a layer and a block column.\\ 
\indent The $\mathbf{0}$ submatrices in $\mathbf{H}$ are defined as \emph{invalid} blocks, where there are no edges between the corresponding CNs and VNs, and the submatrices $\mathbf{I}_s$ as \emph{valid} blocks. In a conventional approach to scheduling, message computation is done for all the valid and invalid blocks. To avoid processing invalid blocks, we propose an alternate representation of $\mathbf{H}_b$ in the form of two matrices: $\boldsymbol\beta_I$, the block index matrix and $\boldsymbol\beta_S$, the block shift matrix which hold the index locations (column number of each block in a row or layer) and the shift values (defining the connections between the CNs and VNs) corresponding to \emph{only} the valid blocks in $\mathbf{H}_b$, respectively.
\begin{fact}
In the decoder architecture, CN and VN processing is performed by a single processing unit termed as the Node Processing Unit (NPU). The NPU is further split into two units namely, the Local NPU (LNPU) and the Global NPU (GNPU) to reduce the decoding complexity \cite{arch_main}.
\end{fact}
A naive way to schedule the layer-level processing is shown in Fig. \ref{fig:swppl}(a). The outer for-loop executes $I$ times, processing node metrics over all the layers. In the first inner for-loop, the GNPU output is first computed over the $J$ blocks in each layer, as per the algorithm in \cite{arch_main} and is then fed to the second for-loop where the LNPU produces the respective metrics for the same set of blocks. We call this the $1x$ or the Baseline architecture. It is evident that one of the NPU idles while the other processes. To avoid idling, we use Fact 1 and process the GNPU and LNPU in a pipeline as shown in Fig. \ref{fig:swppl}. This version is called as the $2x$ or the Pipelined architecture. We would like to emphasize here that, pipelining was described in software at the algorithmic description level and not the HDL level. The algorithmic compiler translated the high-level description to an HDL description for the case study decoder implementation in a little over two minutes.

\begin{remark}
Fig. \ref{fig:swppl} (b) shows upto 6 layers ($L_1$ to $L_6$) in the pipeline. From the bound on the number of layers one can pipeline in this manner is derived in \cite{arch_main}, for the QC-LDPC code in this case study the maximum number is 6.
\end{remark}

\section{Multi-core Decoder}
\label{sec:sixcores}
\begin{sidewaysfigure*}[htp]
\centering
\includegraphics[width=\textwidth]{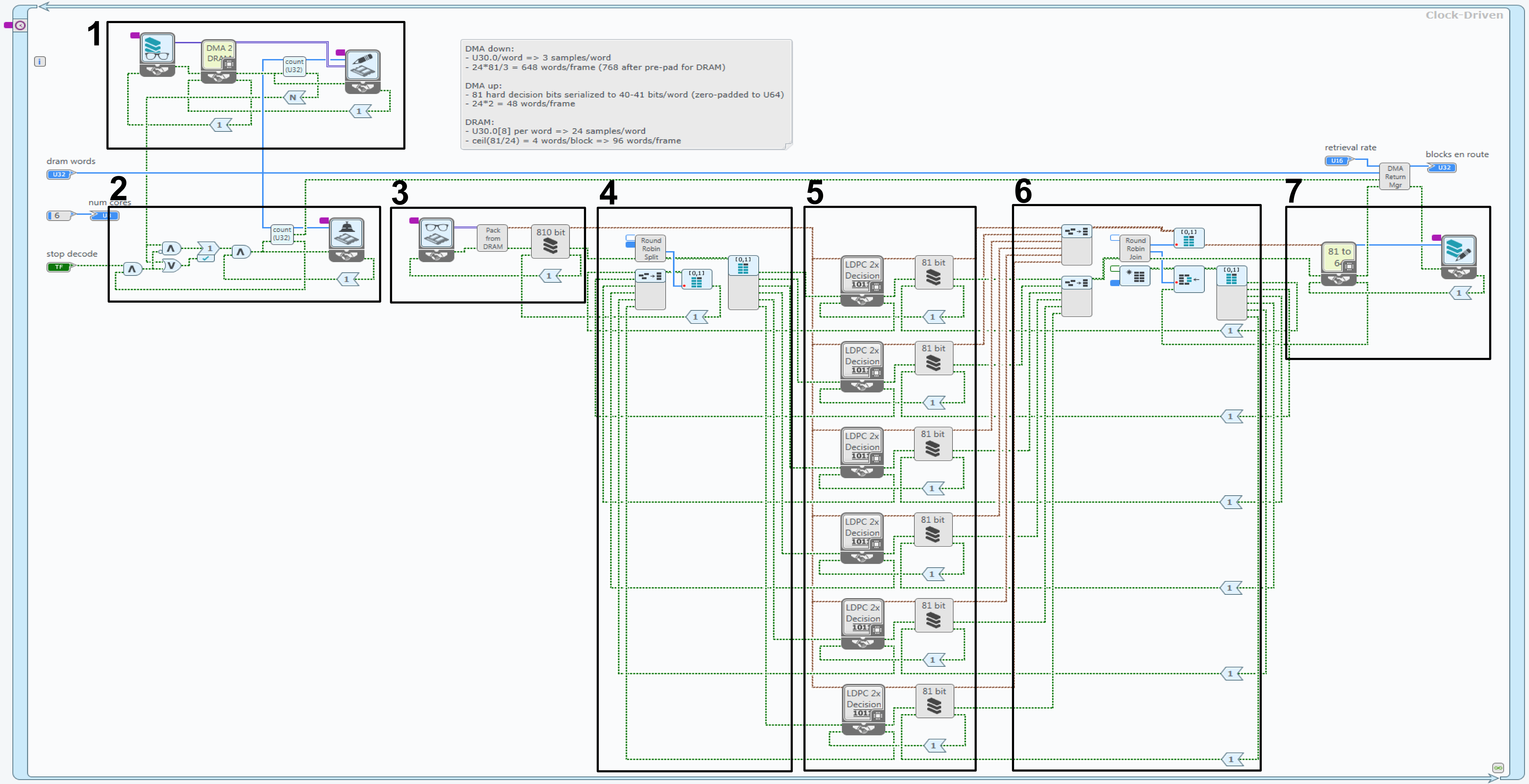}
\caption{Top-level VI describing the parallelization of the QC-LDPC decoder \cite{arch_main} on the \emph{NI USRP-2953R} containing the Xilinx \emph{Kintex7 (410t)} FPGA.}
\label{fig:6core_vi}
\end{sidewaysfigure*}

The decoder implementation based on the Pipelined ($2x$) architecture that achieves a throughput of 420Mb/s (at 200MHz) is hereafter referred to as a \emph{core}. The core operates for $m_b \times n_b = 12 \times 24$, $z=27$, $54$ and $81$ resulting in code lengths of $n=24 \times z=648$, $1296$ and $1944$ bits respectively and a code rate $R=\frac{1}{2}$.  It is worthwhile to note that, for the Pipelined version of the decoder, pipelining was fully described in software. Moreover, the algorithm was described in a high-level language - graphical code in \emph{\emph{LabVIEW\texttrademark}} \,(i.e. not in a hardware description language). The algorithmic compiler in \emph{LabVIEW\texttrademark} \,CSDS\texttrademark \,translated the high-level description into a VHDL description. \\
\indent On account of the scalability and reconfigurability of the decoder architecture in \cite{arch_main}, it is possible to achieve high throughput by employing multiple decoder cores in parallel. Fig. \ref{fig:6core_vi} shows the top-level multi-core decoder virtual instrument (VI), where 6 cores are deployed on a single \emph{Xilinx Kintex7 FPGA (410t)}. The high-level operation of the decoder is described in the steps below (corresponding to the highlighted sections in Fig. \ref{fig:6core_vi}):
\begin{enumerate}
\item Serial stream of the encoded data is read as frames from the host-to-target Direct Memory Access (DMA) mechanism. Here, host may be an arbitrary processing platform such as a PC or a real-time controller and target is the \emph{Xilinx Kintex7 FPGA (410t)} on the \emph{NI USRP-2953R}. This data is subsequently stored in the Dynamic Random Access Memory (DRAM). 
\item Request frames from the DRAM.
\item Read and buffer frames from the DRAM.
\item Distribute incoming frames to the cores in a round-robin manner.
\item Perform decoding with fixed-latency, parallel processing of frames staggered with respect to time. Buffer the decoded frames.
\item Collect the decoded frames and serialize them with respect to the round-robin manner used in step (3).
\item Write frames to the target-to-host DMA mechanism. \\
\end{enumerate}

\section{Results}
\label{sec:results}
\begin{figure}
\centering
\includegraphics[scale=0.5]{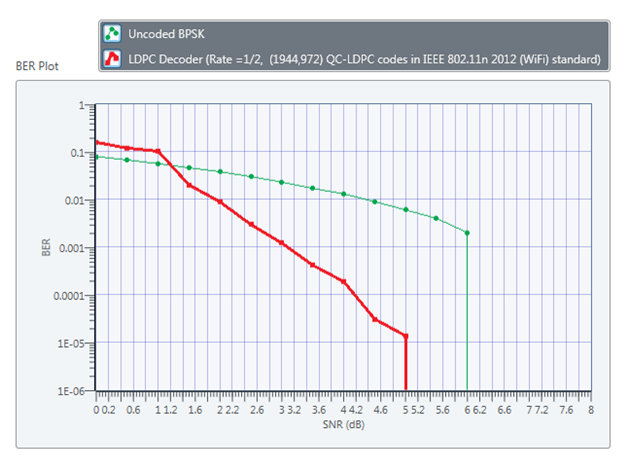}
\caption{Bit Error Rate (BER) performance comparison between uncoded BPSK (green) and the 2.48Gb/s, rate=1/2, QC-LDPC decoder (red) on the \emph{NI USRP-2953R} containing the Xilinx \emph{Kintex7 (410t)} FPGA.}
\label{fig:ber}
\end{figure}

The performance and resource utilization of the Baseline and the Pipelined version is compared in Table \ref{tab:pplperf}. The resources consumed by the Pipelined decoder are almost the same as that of the Baseline decoder, in spite of the $1.5x$ increase in throughput performance. The 2.48Gb/s decoder was developed in stages, where at each stage a core was added (except for stage 3) and the performance and resource figures were recorded. The results of each stage are compared in Table \ref{tab:multicoreperf}. The Bit Error Rate (BER) performance of the 2.48Gb/s version (with 6 cores) is shown in Fig. \ref{fig:ber}.\\
\indent We have successfully demonstrated this work in \emph{IEEE GLOBECOM'14} \cite{globecomm} where an overall throughput of 2.06Gb/s was achieved by using five decoder cores in parallel on the \emph{Xilinx Kintex7 (410t)} FPGA in the \emph{NI USRP-2953R}.

\begin{table}[htp]
\centering
\setlength{\tabcolsep}{9.5pt}
\scalebox{0.85}{
\begin{tabular}{l@{\hspace{4pt}}*{3}{c}}
\midrule
& \bf{Baseline} &\bf{Pipelined} \\
\midrule 
\bfseries Throughput (Mb/s)
&290 &420	 \\
\midrule
\bfseries Clock Rate (MHz) 
&200 &200 \\
\midrule
\bfseries Time to generate VHDL (min)
&2.02 &2.08 \\
\midrule
\bfseries Total Compile Time (min)
&$\approx 36$ &$\approx 36$ \\
\midrule
\bfseries Total Slice (\%)
&26 &28 \\
\midrule
\bfseries LUT (\%)
&16 &18 \\
\midrule
\bfseries FF (\%)
&9 &10 \\
\midrule
\bfseries DSP (\%)
&5 &5 \\
\midrule
\bfseries BRAM (\%)
&11 &11 \\
\bottomrule
\addlinespace \addlinespace
\end{tabular}
}
\caption{Performance and resource utilization comparison for the Baseline architecture with the Pipelined architecture of the QC-LDPC decoder on the \emph{NI USRP-2953R} containing the Xilinx \emph{Kintex7 (410t)} FPGA.}
\label{tab:pplperf}
\end{table}

\begin{table}[htb]
\centering
\setlength{\tabcolsep}{5pt}
\scalebox{0.85}{
\begin{tabular}{l@{\hspace{4pt}}*{6}{c}}
\midrule
\bfseries Cores
&\bf{1} &\bf{2} &\bf{4} &\bf{5} &\bf{6} \\
\midrule 
\bfseries Throughput (Mb/s)
&420 &830 &1650 &2060 &2476 \\
\midrule
\bfseries Clock Rate (MHz) 
&200 &200 &200 &200 &200 \\
\midrule
\bfseries Time to VHDL (min)
&2.08 &2.08 &2.08 &2.02 &2.04 \\
\midrule
\bfseries Total Compile (min)
&$\approx 36$ &$\approx 60$ &$\approx 104$ &$\approx 132$ &$\approx 145$ \\
\midrule
\bfseries Total Slice (\%)
&28 &44 &77 &85 &97 \\
\midrule
\bfseries LUT (\%)
&18 &28 &51 &62 &73 \\
\midrule
\bfseries FF (\%)
&10 &16 &28 &33 &39 \\
\midrule
\bfseries DSP (\%)
&5 &11 &21 &26 &32 \\
\midrule
\bfseries BRAM (\%)
&11 &18 &31 &38 &44 \\
\bottomrule
\addlinespace \addlinespace
\end{tabular}
}
\caption{Performance and resource utilization comparison for versions with varying number of cores of the QC-LDPC decoder implemented on the \emph{NI USRP-2953R} containing the Xilinx \emph{Kintex7 (410t)} FPGA.}
\label{tab:multicoreperf}
\end{table}

\section{Conclusion}
\label{sec:conc}
This work validates the scalability of our decoder architecture in \cite{arch_main} by deploying multiple decoder cores in parallel. The development was done using an algorithmic compiler that translated the high-level description of the decoding algorithm into an HDL in approximately 2 minutes. The standalone standard compliant decoder achieves an overall throughput of 2.48Gb/s at an operating frequency of 200MHz on the \emph{Xilinx Kintex-7} FPGA in the \emph{NI USRP-2953R}. With little or no modification this decoder can be applied to a large family of standard compliant QC-LDPC codes such as those specified in IEEE 802.16e and Digital Video Broadcast (DVB).\\

\section*{Acknowledgment}
The authors would like to thank the Department of Electrical \& Computer Engineering, Rutgers University for their continual support for this research work and the \emph{LabVIEW\texttrademark} \, FPGA R\&D and the Advanced Wireless Research team in National Instruments for their valuable feedback and support.

\bibliographystyle{IEEEtran}
\bibliography{IEEEabrv,dec_impl_bib}

\end{document}